\documentclass{webofc}
\usepackage[varg]{txfonts}   
\usepackage{caption}
\usepackage{subcaption}
\usepackage{comment}
\usepackage{lineno}
\begin{document}
\title{Exploring Electromagnetic Field Effects and Constraining Transport Parameters of QGP using STAR BES-II data}
%
%

\author{\firstname{Aditya Prasad Dash} \lastname{}\inst{1}\fnsep\thanks{\email{aditya55@physics.ucla.edu}}
        \firstname{} \lastname{for the STAR Collaboration}
}

\institute{Department of Physics and Astronomy, University of California Los Angeles, Los Angeles, CA 90095 
          }

\abstract{%
Heavy-ion collisions undergo various stages in their evolution and it is crucial to disentangle the initial- and final-stage effects. In this work, we report measurements of two types of observables: (i)
charge-dependent directed flow ($\Delta v_1$), which is sensitive to the initial
ultra-strong electromagnetic fields, and (ii) flow correlations, such as
$r_n (\eta)$ which is sensitive to the initial longitudinal de-correlation, and
correlations among flow harmonics. These measurements contribute to constraining the initial state of the collisions, and
through a comprehensive beam energy scan, we gain significant insights into the system evolution in the presence of initial spatial asymmetry and electromagnetic fields.
}

\maketitle
\section{Introduction}
Ultra-strong magnetic fields on the order of $10^{18}$ Gauss, 
the strongest magnetic fields observed in nature, are expected to be produced in the very early stages of heavy-ion collisions at RHIC energies ~\cite{Gursoy18,Gursoy14}. The magnetic fields are mostly created by spectators and decay very fast, with time scales comparable to the passage time of the colliding nuclei~\cite{Gursoy18,Gursoy14}. However, the decay of the fields can be compensated by the Faraday induction effect, which depends on the medium properties (such as electrical conductivity) and on the formation time of the quarks. Additionally, the study of the formation and decay of the initial electromagnetic fields is essential for understanding the evolution of the quark-gluon plasma (QGP) in the presence of electromagnetic (EM) fields. 

The initial state in heavy-ion collisions could have significant longitudinal de-correlation, leading to a difference between event planes reconstructed at different pseudorapidity ranges~\cite{CMS15,Jia14}. 
Moreover, the initial-state geometry and asymmetries in energy deposition could evolve into final-state flow harmonics and event-plane angular correlations, the study of which can be used to constrain various initial-state models and understand the mechanisms of energy deposition by the colliding nuclei. 
\label{intro}

\section{Experimental Results}
\subsection{EM-field Effects on Directed Flow}
Quarks carry electric charges and experience various electromagnetic forces. In particular, the Lorentz force due to the Hall effect, $\textbf{F}=q\textbf{v} \times \textbf{B}$, acts in the opposite direction to the forces due to the Faraday induction effect (generated by the decaying magnetic field) and the Coulomb effect (the electric field due to spectators). As electromagnetic forces are proportional to charge, the difference in directed flow slope between positively and negatively charged particles ($\Delta dv_1/dy$) can serve as a probe to the EM-field effects in heavy-ion collisions~\cite{Gursoy18,Gursoy14}. The Hall effect should impart a positive $\Delta dv_1/dy$, while the Faraday+Coulomb effect should give a negative $\Delta dv_1/dy$ to charged particles.

Furthermore, quarks transported from the colliding nuclei to midrapidities can have a different directed flow than those produced in pairs. 
Model studies~\cite{TransportedQuarksModels1,TransportedQuarksModels2,TransportedQuarksModels3} show that transported-quark 
effect  could contribute a positive $\Delta dv_1/dy$ for protons and kaons, and a negative $\Delta dv_1/dy$ for pions due to their quark contents~\cite{STAR23}. 

\begin{figure}[tb]
\centering
\includegraphics[width=\textwidth]{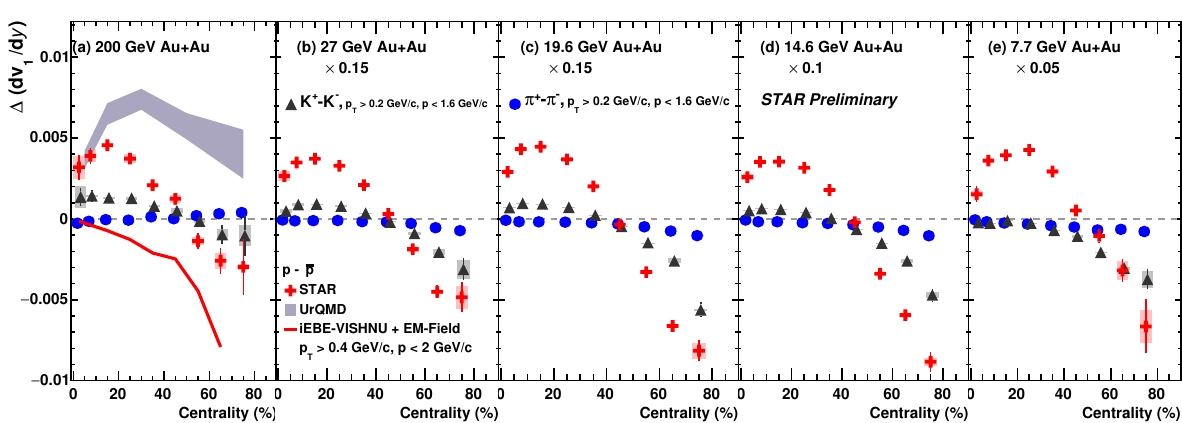}
\caption{$\Delta (d v_1/dy)$ as a function of centrality for pions, kaons, and protons in Au+Au collisions at $\sqrt{s_{NN}}=$ 200, 27, 19.6, 14.6, and 7.7 GeV~\cite{STAR23}. At 200 GeV, model calculations for protons from UrQMD and iEBE-VISHNU are also added for comparison.}
\label{figure1}
\vspace{-0.5cm}
\end{figure}

Figure~\ref{figure1} shows the STAR measurements of $\Delta dv_1/dy$ as a function of centrality for $\pi^{\pm}$,$K^{\pm}$ and $p,\bar{p}$ in Au+Au collisions at $\sqrt{s_{NN}}=$ 200--7.7 GeV~\cite{STARDetector03,STAR23}. $\Delta dv_1/dy$ becomes negative in peripheral collisions (except for pions at 200 GeV), which is expected from the dominance of the Faraday+Coulomb effect. Other mechanisms for the centrality dependence of $\Delta dv_1/dy$ are under investigation~\cite{STAR23Hall,BaryonDiffusion}. Data at $\sqrt{s_{NN}}=$ 200 GeV are comparable to the IEBE-VISHNU+EM field calculations with conductivity $\sigma = 0.023$ fm$^{-1}$ from lattice QCD~\cite{Gursoy18,Gursoy14}. The increase in the magnitude of $\Delta dv_1/dy$ in peripheral collisions with decrease in beam energy, observed in data, is expected from the longer passage times ($2\text{R}/\gamma$) and shorter lifetimes of the fireball (due to which the late-time dilution of the $v_1$ splitting is smaller) at lower collision energies~\cite{Gursoy18,Gursoy14}.

\begin{figure}[tb]
\centering
\includegraphics[width=\textwidth]{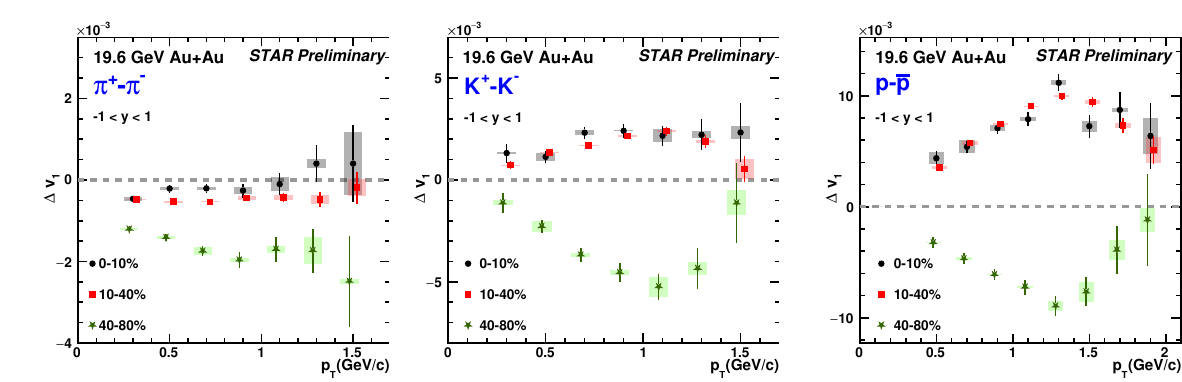}
\caption{$\Delta v_1$($p_T$) for pions, kaons, and protons in Au+Au collisions at $\sqrt{s_{NN}}=19.6$ GeV.}
\label{fig2:deltav1vspt19GeV}
\vspace{-0.5cm}
\end{figure}
Figure~\ref{fig2:deltav1vspt19GeV} shows $\Delta v_1$ ($p_T$) for pions, kaons, and protons in Au+Au collisions at $\sqrt{s_{NN}}=19.6$ GeV. The negative $\Delta v_1$ and the increase in $|\Delta v_1|$ with $p_T$ are expected from theory calculations~\cite{Gursoy18,Gursoy14} and could arise from a decrease in the Hall effect at higher $p_T$ (or smaller $p_z$). Similar observations hold at $\sqrt{s_{NN}}=$ 14.6 and 7.7 GeV.

\subsection{Longitudinal De-correlation}

\begin{figure}[tb]
\centering
\includegraphics[width=0.7\textwidth]{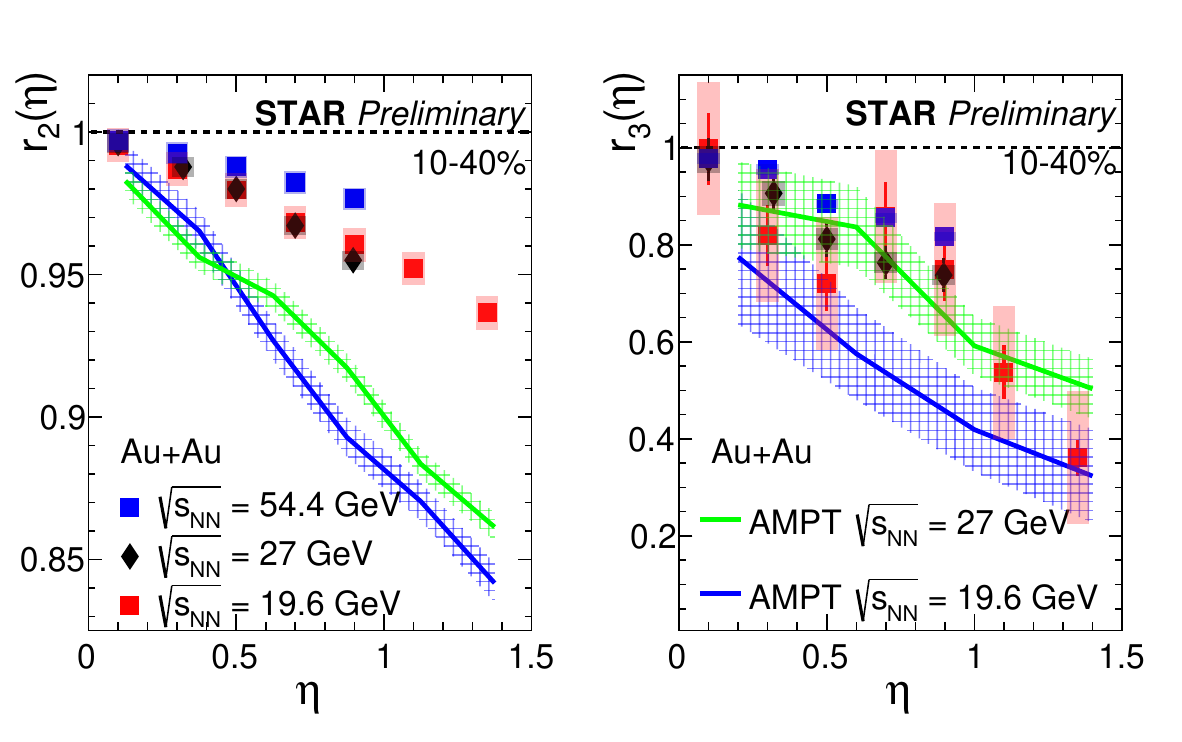}
\caption{$r_2(\eta)$ and $r_3(\eta)$ in 10--40\% centrality Au+Au collisions at $\sqrt{s_{NN}}=$ 19.6--54.4 GeV. 2.5$<\eta_{\text{ref}}<4.0$ (2.1$<|\eta_{\text{ref}}|<5.1$) is selected for 54.4 GeV (27 and 19.6 GeV).} 
\label{fig3:longitudinaldecorrelation}
\vspace{-0.5cm}
\end{figure}

The $r_n (\eta)$ observable~\cite{CMS15,Jia14} is sensitive to the de-correlation between event planes reconstructed from the pseudorapidity ranges $\eta$ and $-\eta$, respectively. $r_n(\eta)$ is expected to be 1 when there are no de-correlation or nonflow effects. Figure~\ref{fig3:longitudinaldecorrelation} shows the measurements of $r_2(\eta)$ and $r_3(\eta)$ in the 10--40\% centrality range, significantly deviating from unity at RHIC energies. 
AMPT~\cite{Dixit23} shows stronger deviation for $r_2(\eta)$ and comparable deviation for $r_3(\eta)$ compared to  experimental data.

\subsection{Constraining Initial State Using Correlations}

\begin{figure}[tb]
\centering
\includegraphics[width=0.8\textwidth]{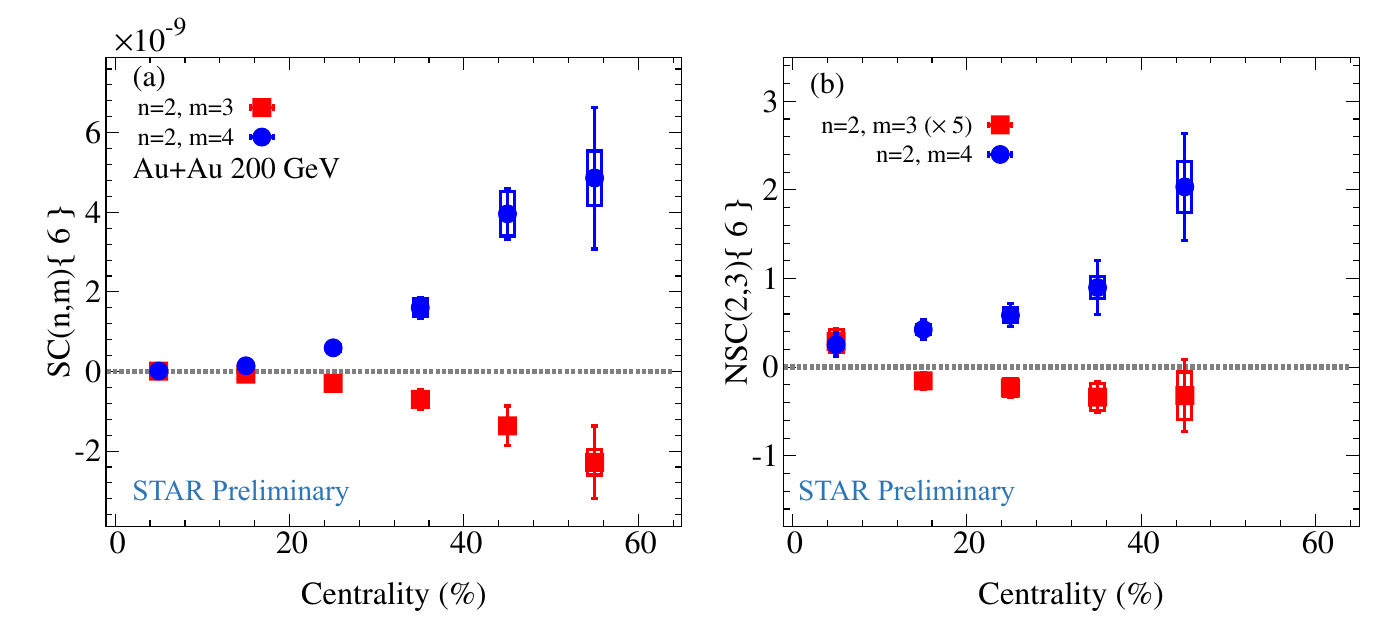}
\caption{Six-particle (normalized) cumulants as a function of centrality for Au+Au collisions at $\sqrt{s_{NN}}=200$ GeV.} 
\label{fig4:cumulants}
\vspace{-0.5cm}
\end{figure}

Symmetric Cumulants (SC) are expected to be sensitive to the interplay between initial- and final-state effects. In contrast, Normalized Symmetric Cumulants (NSC) are predicted to be dominated by initial-state effects. Thus, simultaneous measurements of SC and NSC can constrain the initial- and final-state effects~\cite{Magdy23}. The six-particle SC and NSC are given as
\begin{equation}
SC(n.m)\{6\}=\langle 6 \rangle_{nnm}-\langle 4 \rangle_{nn} \langle 2 \rangle_{m}-2 \langle 4 \rangle_{nm} \langle 2 \rangle_{n} +2\langle 2 \rangle_{n}^2 \langle 2 \rangle_{m}, 
\end{equation}
\begin{equation}
NSC(n.m)\{6\}=\frac{SC(n,m)\{6\}}{\langle 2 \rangle_{n}^{sub} \langle 2 \rangle_{n}^{sub} \langle 2 \rangle_{m}^{sub}}.
\end{equation}

Figure~\ref{fig4:cumulants} shows the six-particle (normalized) symmetric cumulants as a function of centrality in Au+Au collisions at $\sqrt{s_{NN}}=200$ GeV~\cite{Bilandzic20, Magdy23}.
Anti-correlation between $v_2$ and $v_3$ is observed which is expected from the anti-correlation between $\epsilon_2$ and $\epsilon_3$. A positive correlation between $v_2$ and $v_4$ is consistent with mode coupling between $v_2$ and $v_4$. These measurements can help constrain various initial-state models.

\section{Summary}
We have presented measurements of $\Delta v_1$, $r_n(\eta)$, and flow harmonic correlations in Au+Au collisions at $\sqrt{s_{NN}}=7.7-200$ GeV in the Beam Energy Scan phase-II program of the STAR experiment. The $\Delta v_1$ results are compatible with models using strong electromagnetic fields and conductivity from lattice QCD. Proton $\Delta dv_1/dy$ is negative in peripheral collisions, supporting the dominance of the Faraday$+$Coulomb effect, and becomes more negative at lower collision energies, which is expected from the corresponding longer lifetime of the electromagnetic field and shorter lifetime of the fireball. $r_2(\eta)$ and $r_3(\eta)$ significantly deviate from unity at RHIC energies and show beam energy dependence. From the six-particle (normalized) symmetric cumulant measurements, anti-correlation is observed between $v_2$ and $v_3$, and positive correlation is observed between $v_2$ and $v_4$.

%
%

\end{document}